 \documentstyle[11pt]{article} 
\begin{document}
\begin{titlepage}
\title{Critique of proposed limit to space--time measurement, based on
Wigner's clocks and mirrors}
\author{L. Di\'osi and B. Luk\'acs
\thanks{E-mail: diosi@rmki.kfki.hu, lukacs@rmki.kfki.hu}\\
KFKI Research Institute for Particle and Nuclear Physics\\
H-1525 Budapest 114, POB 49, Hungary\\\\
{\it bulletin board ref.: quant-ph/9501001}}
\date{January 3, 1995}
\maketitle
\begin{abstract}
Based on a relation between inertial time intervals and the Riemannian
curvature, we show that space--time uncertainty derived by Ng and
van Dam implies absurd uncertainties of the Riemannian curvature.
\end{abstract}
\end{titlepage}

Recently, Ng and van Dam \cite{NgDam,NgDamrep}
presented a proof of the intrinsic quantum
uncertainty $\delta\ell$ of {\it any} geodetic length $\ell$ being
proportional to the one third power of the length itself:
\begin{equation}
\delta\ell=\ell_P^{2/3}\ell^{1/3}
\end{equation}
where $\ell_P$ is the Planck lenghts.
In addition, they claim that an intrinsic uncertainty of space--time
metric has been derived in Refs.~\cite{NgDam,NgDamrep}. Now, the problem
deserves a discussion since,
a few years ago, the present
authors \cite{DioLuk}
pointed out that the formula (1) would certainly overestimate
the uncertainty of the space--time. This formula {\it would be}
the uncertainty of a distincted world line whose length is measured
at the price of total ignorance about the lenghts of any other neighbouring
world lines. In a sense, the uncertainties of all neighbouring world lines
within about a tube of diameter $\ell$ will charge the uncertainty of the
distincted one.

Calculate, for instance, the mass $m$ of the clock
when adjusted according to the Eqs.~(3) and (4) of Ng and van Dam:
\begin{equation}
m=m_P\left({\ell\over\ell_P}\right)^{1/3}
\end{equation}
where $m_P$ is the Planck-mass.
It is at least worrying that the optimum mesurement of a lenght
$\ell\approx 1cm$
requires a clock of mass $m\approx 10^6g$ and, similarily, the optimum
measurement of a timelike distance $t\approx1s$ needs a clock with
$m\approx 10^{16}g$ (i.e. $10^{10}$ metric tons!).
Of course, the large mass of the clock needed to reach the limit of accuracy
is not a proof against the proposed fluctuation formula. But we can show
that eq. (1) leads to drastic effects in the space--time continuum, strongly
affecting macroscopy.
That Eq.~(1) seriuosly overestimates the uncertainty
of space--time can now be shown by an independent elementary proof.

Let us start with the formula (6) of Ng and van Dam:
\begin{equation}
\delta t = t_P^{2/3}t^{1/3}
\end{equation}
where $\delta t$ is the proposed uncertainty of the time $t$ along
an arbitrarily chosen time-like geodesic and $t_P$ is the Planck-time.
This uncertainty implies a certain uncertainty of the physical
space--time geometry. One expects that the corresponding fluctuations
of the local Riemann curvature are fairly small.

Fortunately, there exists a simple relation between a subtle triplet of time
intervals on one hand and the average Riemannian curvature on the other.
We recapitulate this relation according to Wigner \cite{Wig}.

Assume space--time is flat on average.
Take a clock and in distance $\ell/2$ a mirror; for simplicity's sake
let them be at rest relative to each other. Let us emit a light signal
from the clock to the mirror, and let the clock measure the total flight
time $t_1$ as the signal has got back to it. Repeat the same experiment
immediately after, for the flight time $t_2$, and similarily for a third one
$t_3$. Then, the average curvature $C$ in the space--time region swept by
the light pulses is
\begin{equation}
C={1\over11c}{t_1-2t_2+t_3\over t_2^2}.
\end{equation}

Let us obtain the quantum uncertainty $\delta C$ of the above curvature.
Of course, each period $t_i$ $(i=1,2,3)$ has the same average value
$\ell/c$. Their quantum uncertainties $\delta t_i$ are also equal.
According to Ng and van Dam, any timelike geodesic length possesses the
ultimate uncertainty (3) so do ours, too:
\begin{equation}
\delta t_i
= \left({\ell_P\over c}\right)^{2/3}\left({\ell\over c}\right)^{1/3}.
\end{equation}

If $\ell\gg\ell_P$ the periods $t_i$ are much larger than their fluctuations
(5) and, consequently, we can approximate the uncertainty of
the curvature (4) by an expession linear in $\delta t_i$:
\begin{equation}
\delta C={c\over11\ell^2}\delta \left(t_1-2t_2+t_3\right).
\end{equation}
To calculate the squared average value of $\delta C$, one rewrites the
above equation in the following equivalent form:
\begin{eqnarray}
[\delta C]^2=\left({c\over11\ell^2}\right)^2
\Bigl(3[\delta t_1]^2+9[\delta t_2]^2+3[\delta t_3]^2\nonumber\\
-3[\delta(t_1+t_2)]^2-3[\delta(t_2+t_3)]^2+[\delta(t_1+t_2+t_3)]^2\Bigr).
\end{eqnarray}
Each term on the RHS is then evaluated by means of the Eq.~(3). After
extracting a root, on obtains
\begin{equation}
\delta C ={\sqrt{15-6\times2^{2/3}+3^{2/3}}\over11}
{1\over\ell}\left({\ell_P\over\ell}\right)^{2/3}.
\end{equation}
The averaged Riemann-tensor components are related to the
averaged curvatures (4) as, e.g., $\overline R_{0101}=2C^2$ provided
both clock and mirror lay along the first coordinate axis [4].
It seems plausible to assume that $\delta C$ of (8)
yields the order of magnitude not only for the Riemann-tensor components
but for the components of Ricci-tensor as well as for the Riemann-scalar
$\overline R$ unless special statistical correlation is shown or at least
assumed  between the various components of the Riemann-tensor.
So, Eq.~(8) yields the following estimation for the
Riemann-scalar ${\overline R}$ averaged in a 4-volume $\sim\ell^4/c$:
\begin{equation}
{\delta\overline R}
\sim {1\over\ell^2}\left({\ell_P\over\ell}\right)^{4/3}.
\end{equation}

Basically, one would expect with Ng and van Dam that these fluctuations are
{\it small}. There is at least one good criterion to test their smallness.
According to the Einstein theory of general relativity, nonzero scalar
curvature $R$ assumes nonzero energy density. If we assume that the
energy-momentum tensor is dominated by the energy density $\rho$ then
the fluctuation (9) of the Riemann scalar would imply
\begin{equation}
\delta{\overline\rho}
\sim (c^2/G){\overline R} \sim (\hbar/c)\ell_P^{-2/3}\ell^{-10/3},
\end{equation}
where $\delta{\overline\rho}$ denotes the universal fluctuation of the
energy density $\rho$ averaged in a 4-volume $\sim\ell^4/c$. This
fluctuation would be {\it extremly high} at small length scales.
At $\ell\sim 10^{-5}cm$, for instance, the uncertainty
$\delta{\overline\rho}$ would be in the order of water density; that is
trivially excluded by experience.
According to decent cosmological estimations, e.g. from galaxy counts,
the average mass density of
our Universe should not exceed $10^{-29}gcm^{-3}$. Then, the Eq.~(10)
yields $\ell\gg 10^4cm$ which in turn means that the proposal of
Ng and Dam for the uncertainty of geodesic length may not be applied
for lenghts shorter than some $100$ meters otherwise one might get another
Universe due to the additional cosmologic mass density generated
by the short range metric fluctuations.

According to all these arguments, we think
that Ng and van Dam in \cite{NgDam,NgDamrep} have in fact
derived an {\it unconditional} uncertainty for a {\it single} geodesic.
However, the uncertainty of a single
geodesic length should not be used to calculate the intrinsic uncertainty
of the space--time metric: it would need the simultaneous
uncertainties of {\it all} geodesics or at least of a subtle subset of all.
We pointed out that to ignore the correlations of those uncertainties
would lead too high uncertainties of the space--time curvature.
Finally, it is worth to mention that a detailed account of the present
authors alternative to replace Eq.~(1) can be found in Ref.~\cite{DioLuk}.

\bigskip

This work was supported by the grant OTKA No. 1822/1991.

\end{document}